\documentclass[twocolumn,showpacs,preprintnumbers,amsmath,amssymb]{revtex4}

\usepackage{graphicx}
\usepackage{dcolumn}
\usepackage{bm}

\begin{document}

\preprint{EKM-TP3/23-01}

\title{Derivation of the Curie-Weiss Law in Dynamical Mean-Field Theory
}

\author{Krzysztof Byczuk$^{a,b}$} \email{kbyczuk@physik.augsburg-uni.de}
\author{Dieter Vollhardt$^a$}%
\affiliation{%
(a) Theoretical Physics III, Centre for Electronic Correlations and
Magnetism, 
Institute for Physics, University of Augsburg,
D-86135 Augsburg, Germany\\
(b) Institute of Theoretical Physics,
Warsaw University, ul. Ho\.za 69, 00-681 Warszawa, Poland
}%


\date{\today}

\begin{abstract}
We present an analytic derivation of the linear temperature dependence of
the inverse static susceptibility $\chi ^{-1}(T,U)\sim T-T{_{c}}(U)$ near
the transition from a paramagnetic to a ferromagnetic correlated metal
within the dynamical mean-field theory (DMFT) for the Hubbard model. The
equations for the 
critical temperature and interaction strength of the transition are also
determined.

\end{abstract}

\maketitle

\section{Introduction}

The dynamical mean-field theory (DMFT) is a non-perturbative and
thermodynamically consistent approximation scheme for quantum mechanical
many-body problems on a
 lattice (for reviews see Ref. \cite{georges96,pruschke95,voll93})
which becomes exact in the limit of large
coordination numbers \cite{metzner89}. 
In contrast to the static 
 Hartree-Fock mean-field theory the
dynamics of the quantum mechanical correlation problem is fully included in
the DMFT. In the last few years the DMFT proved to be a powerful tool for
the investigation of fermionic lattice models with local Coulomb interaction
such as the Hubbard model and the periodic Anderson model
 \cite{pruschke95,georges96}. 
It is
particularly useful in the case of intermediate-coupling problems such as
the Mott-Hubbard metal-insulator-transition or itinerant ferromagnetism
where perturbative techniques fail (for reviews see Refs. 
 \cite{georges96,fazekas99,voll00}).
In the DMFT the lattice problem is
mapped onto an effective single-site problem whose self-energy $\Sigma
(\omega )$ and Green function $G(\omega )$ have to be calculated
self-consistently with the ${\bf k}$-integrated Dyson equation. The theory
is therefore purely local, i.e., the self-energy $\Sigma (\omega )$ is ${\bf %
k}$-independent and the propagator $G_{{\bf k}}(\omega )=G_{{\bf k}%
}^{0}(\omega -\Sigma (\omega ))$ may be represented by the non-interacting
propagator $G_{{\bf k}}^{0}$ at shifted frequency, at least in the
paramagnetic case. Here the mean-field character of the theory becomes
particularly evident. The local nature of the theory implies that
short-range order in position space is missing.

Numerical solutions of the DMFT equations, in particular by quantum
Monte-Carlo simulations (QMC), revealed that in the case of continuous phase
transitions (e.g., from a paramagnetic metal to an antiferromagnetic
insulator \cite{jarrell}, or to a ferromagnetic metal \cite{ulmke98}) 
the static susceptibility $\chi (T)$
shows a Curie-Weiss behavior above $T_{c}$,
 i.e., $\chi ^{-1}\propto T-T_{c}$,
implying that $\chi \propto \left( T-T_{c}\right) ^{-\gamma }$ diverges with
a critical exponent $\gamma =1$.
 Furthermore, the order parameter was found
to vanish with an exponent $\beta =1/2$ \cite{ulmke98,jarrellbis}.
 In view of the mean-field nature of
the DMFT these numerical findings did not come as a surprise. However,
considering the dynamics of the quantum mechanical problem the situation is
not as self-evident as it may seem. At a continuous phase transition between
two phases in high dimensions, a 
mean-field behavior is naturally expected if the low energy
excitation spectrum of both phases has a gap. In this
case, the fermions may be integrated out, leading to an effective
Ginzburg-Landau-Wilson field theory \cite{vojta}. 
This applies to transitions in the
Heisenberg spin model.
By contrast, if the transition occurs between two {\em metallic} phases,
e.g., from a paramagnetic to a ferromagnetic metal, or between 
a metallic and an insulating 
phase, the result is far from
trivial since the low-lying excitations in the metallic phases 
may couple to the order parameter
and thereby lead to divergences in the effective field theory at $T=0$ or 
even at very 
low temperatures \cite{belitz99}. 
At present,
the consequences of this feedback are still not entirely understood. It is
therefore worthwhile to further investigate the transition between two
metallic phases, also within the DMFT, using {\em analytical} methods.

Curie-Weiss behavior of the magnetic susceptibility is traditionally
associated with {\em localized} magnetic moments, 
and, indeed, is readily
obtained for Heisenberg-type spin models in a mean-field approximations.
 Nevertheless, it is also a
characteristic of interacting itinerant electrons as described, for example,
by the Hubbard model. In particular, a Curie-Weiss behavior
 may be obtained within the
Hartree-Fock approximation which yields $\chi _{HF}^{-1}(T)\sim
T^{2}-T_{c}^{2}$ above $T_{c}$ \cite{wolf,fazekas99}.
 However, since this result is derived for
interaction strengths where the Hartree-Fock approximation is not controlled
by a perturbation theory, its qualitative and quantitative 
validity is questionable \cite{janis}.
The same criticism applies to 
the Stoner criterion for the onset of ferromagnetism \cite{wahle98}.

To calculate the
static magnetic susceptibility in the vicinity of a continuous phase
transition from the paramagnetic to the ferromagnetic metallic state where
electronic correlations are explicitly included we will use the one-band
Hubbard model 
\begin{equation}
H=\sum_{ij,\sigma }t_{ij}c_{i\sigma }^{\dagger }c_{j\sigma
}+U\sum_{i}n_{i\uparrow }n_{i\downarrow },
\end{equation}%
on an arbitrary lattice and employ the DMFT. In particular, we will show
that the critical exponent is indeed $\gamma =1$.

\section{Derivation of the susceptibility}

We wish to calculate the magnetization density 
\begin{equation}
m=(1/2\beta )\sum_{n}\sigma G_{\sigma n},
\end{equation}%
where the local Green function $G_{\sigma n}$ in DMFT is given by the bare
density of states $N^{0}(\epsilon )$ and the local self-energy $\Sigma
_{\sigma n}$ as 
\begin{equation}
G_{\sigma n}=\int d\epsilon \frac{N^{0}(\epsilon )}{i\omega _{n}+\mu +\sigma
h-\Sigma _{\sigma n}-\epsilon }.  \label{1}
\end{equation}%
Here the subscript $n$ refers to the Matsubara frequency $i\omega
_{n}=i(2n+1)\pi /\beta $ for the temperature $T$, with $\beta =1/k_{B}T$,
and $h$ is the external magnetic field in energy units. Within the DMFT the
local Green function $G_{\sigma n}$ is determined self-consistently through (%
\ref{1}) and 
\begin{equation}
G_{\sigma n}=-\frac{\int D\left[ c_{\sigma },c_{\sigma }^{\star }\right]
c_{\sigma n}c_{\sigma n}^{\star }e^{A\{c_{\sigma },c_{\sigma }^{\star },%
{\cal G}_{\sigma }^{-1}\}}}{\int D\left[ c_{\sigma },c_{\sigma }^{\star }%
\right] e^{A\{c_{\sigma },c_{\sigma }^{\star },{\cal G}_{\sigma }^{-1}\}}},
\label{2}
\end{equation}%
by the {\bf k}-integrated Dyson equation
\begin{equation}
{\cal G}_{\sigma n}^{-1}=G_{\sigma n}^{-1}+\Sigma _{\sigma n}.  \label{3}
\end{equation}%
The single-site action ${\cal A}$ has the form 
\begin{eqnarray}
{\cal A}\{c_{\sigma },c_{\sigma }^{\star },{\cal G}_{\sigma }^{-1}\}=%
\mathrel{\mathop{\sum }\limits_{n,\sigma }}c_{\sigma n}^{\star }{\cal G}%
_{\sigma n}^{-1}c_{\sigma n}-\nonumber \\
U\int_{0}^{\beta }d\tau c_{\sigma }^{\ast
}(\tau )c_{\sigma }(\tau )c_{\bar{\sigma}}^{\ast }(\tau )c_{\bar{\sigma}%
}(\tau ),  \label{2bis}
\end{eqnarray}%
where we used a mixed time/frequency convention for Grassman variables $%
c_{\sigma }$, $c_{\sigma }^{\star }$.

We first separate the self-energy $\Sigma _{\sigma n}$ into its static
(i.e., Hartree-Fock) part and its explicitly dynamical contribution $\tilde{%
\Sigma}_{\sigma n}$ as 
\begin{equation}
\Sigma _{\sigma n}=U\frac{n_{0}}{2}-\sigma Um+\tilde{\Sigma}_{\sigma n}.
\label{4}
\end{equation}%
Here $n_{0}$ is the density of particles. The Hartree-Fock approximation
corresponds to neglecting $\tilde{\Sigma}_{\sigma n}$. Since we are
interested in the behavior of the susceptibility close to a continuous transition,
i.e., in the limits $T\rightarrow T_{c}$, $h\rightarrow 0$, $m\rightarrow 0$, 
 where $T_{c}$ is the, yet unknown, Curie temperature, we write 
\begin{eqnarray}
G_{\sigma n} &=&G_{n}+\delta G_{\sigma n},  \nonumber \\
{\cal G}_{\sigma n} &=&{\cal G}_{n}+\delta {\cal G}_{\sigma n},  \nonumber \\
\tilde{\Sigma}_{\sigma n} &=&\tilde{\Sigma}_{n}+\delta \tilde{\Sigma}
_{\sigma n},  \label{4bis}
\end{eqnarray}%
where $\delta G_{\sigma n}$, $\delta {\cal G}_{\sigma n}$, and $\delta 
\tilde{\Sigma}_{\sigma n}$ are spin-dependent corrections to the respective
paramagnetic (i.e., spin-independent) parts. In the next step we expand (\ref{1}) 
and (\ref{3}) up to first order in these corrections. Using Eqs. (\ref{1}) 
and (\ref{4}) we find 
\begin{equation}
\delta G_{\sigma n}=-[\sigma h+\sigma Um-\delta \tilde{\Sigma}_{\sigma
n}]F_{n},  \label{5}
\end{equation}%
where we introduced the function 
\begin{equation}
F_{n}=\int d\epsilon \frac{N^{0}(\epsilon )}{[i\omega _{n}+\mu -U\frac{n_{0}%
}{2}-\tilde{\Sigma}_{n}-\epsilon ]^{2}}.  \label{f}
\end{equation}%
As in the Weiss molecular field theory, the particles may be interpreted as
moving in an {\em effective} magnetic field 
\begin{equation}
h_{eff}\equiv h+Um-\sigma
\delta \tilde{\Sigma}_{\sigma n}.
\label{star} 
\end{equation}
However, in the case of interacting
electrons considered here this effective field is, in general, found to be
dynamic, i.e., it fluctuates in time due to the local correlations which
lead to an exchange of energy between the particles. Neglecting the
dynamical term $\delta \tilde{\Sigma}_{\sigma n}$ we recover the usual
static mean-field expression for the effective magnetic field. Similarly,
expanding Eq. (\ref{3}) and using Eq. (\ref{4}) we obtain 
\begin{equation}
{\cal G}_{n}^{-1}=G_{n}^{-1}+\tilde{\Sigma}_{n}+U\frac{n_{0}}{2},  \label{7}
\end{equation}%
and 
\begin{equation}
\frac{\delta {\cal G}_{\sigma n}}{{\cal G}_{n}^{2}}=\frac{\delta G_{\sigma n}%
}{G_{n}^{2}}+\sigma Um-\delta \tilde{\Sigma}_{\sigma n}.  \label{8}
\end{equation}

Our goal is to find $\delta \tilde{\Sigma}_{\sigma n}$ and then, with the
help of Eq. (\ref{5}), to calculate the magnetization $m$. Since Eq. (\ref{8}%
) still contains the two unknown functions $\delta \tilde{\Sigma}_{\sigma n}$
and $\delta {\cal G}_{\sigma n}$, an additional condition is required to close
the set of equations. This condition is provided by the fact that the
self-energy is a functional of ${\cal G}_{\sigma n}$ in  perturbation
theory, i.e., $\tilde{\Sigma}_{\sigma n}=\tilde{\Sigma}_{\sigma }\left[ 
{\cal G}_{\sigma n}\right] $, to infinite order. Hence, we can formally
expand as 
\begin{equation}
\tilde{\Sigma}_{\sigma }\left[ {\cal G}_{\sigma n}\right] =\tilde{\Sigma}%
_{\sigma }\left[ {\cal G}_{n}+\delta {\cal G}_{\sigma n}\right] \approx 
\tilde{\Sigma}\left[ {\cal G}_{n}\right] +\sum_{n^{\prime }}\frac{\delta 
\tilde{\Sigma}\left[ {\cal G}_{n}\right] }{\delta {\cal G}_{n^{\prime }}}%
\delta {\cal G}_{\sigma n^{\prime }}.  \label{9}
\end{equation}%
Using Eq.(\ref{7}) relating ${\cal G}$ and $G$ we find the functional
derivative 
\begin{equation}
\frac{\delta \tilde{\Sigma}}{\delta {\cal G}}=\frac{\delta \tilde{\Sigma}}{%
\delta G} \cdot \frac{\delta G}{\delta {\cal G}}=\frac{1}{F}\frac{\delta G}{\delta 
{\cal G}},
\end{equation}%
where $F\equiv F_{n}$ is given by Eq. (\ref{f}). Employing Eq. (\ref{2}) in
the paramagnetic phase and using Eq. (\ref{8}), we obtain the spin-dependent
correction to the self-energy $\tilde{\Sigma}_{\sigma n}$ as 
\begin{widetext}
\begin{equation}
\delta \tilde{\Sigma}_{\sigma n}=\sigma h\sum_{n^{\prime }}M_{nn^{\prime
}}^{-1}\sum_{n^{\prime \prime }}\frac{\Gamma _{n^{\prime }n^{\prime \prime
}}F_{n^{\prime \prime }}}{F_{n^{\prime }}G_{n^{\prime \prime }}^{2}}+\sigma
Um\sum_{n^{\prime }}M_{nn^{\prime }}^{-1}\sum_{n^{\prime \prime }}\frac{%
\Gamma _{n^{\prime }n^{\prime \prime }}}{F_{n^{\prime }}}\left( \frac{%
F_{n^{\prime \prime }}}{G_{n^{\prime \prime }}^{2}}-1\right) ,  \label{11}
\end{equation}%
\end{widetext}
where $M_{nn^{\prime }}^{-1}$ is the inverse of the matrix $M_{nn^{\prime }}$
defined as 
\begin{equation}
M_{nn^{\prime }}=\delta _{nn^{\prime }}+\frac{\Gamma _{nn^{\prime }}}{F_{n}}%
\left( \frac{F_{n^{\prime }}}{G_{n^{\prime }}^{2}}-1\right) ,
\end{equation}%
and 
\begin{equation}
\Gamma _{nn^{\prime }}=\frac{1}{2}\sum_{\sigma \sigma ^{\prime }}\left[
\langle c_{\sigma n}c_{\sigma n}^{\star }c_{\sigma ^{\prime }n^{\prime
}}c_{\sigma ^{\prime }n^{\prime }}^{\star }\rangle -\langle c_{\sigma
n}c_{\sigma n}^{\star }\rangle \langle c_{\sigma ^{\prime }n^{\prime
}}c_{\sigma ^{\prime }n^{\prime }}^{\star }\rangle \right]  \label{12}
\end{equation}%
is the two-particle density-density correlation function calculated in the
paramagnetic phase. We note that in the Hartree-Fock approximation the
two-particle correlations are neglected, i.e. $\Gamma _{nn^{\prime }}^{HF}=0$%
, and therefore $\tilde{\Sigma}_{\sigma n}=0$.

The spin-dependent correction to the local Green function $G_{\sigma n}$ can
now be expressed as 
\begin{equation}
\delta G_{\sigma n}=-\sigma \left[ hH_{n}+Um(H_{n}+\Delta H_n)\right]
F_{n},  \label{13}
\end{equation}%
where 
\begin{equation}
H_{n}\equiv 1-\sum_{n^{\prime }}M_{nn^{\prime }}^{-1}\sum_{n^{\prime \prime
}}\frac{\Gamma _{n^{\prime }n^{\prime \prime }}F_{n^{\prime \prime }}}{%
F_{n^{\prime }}G_{n^{\prime \prime }}^{2}},
\label{20}
\end{equation}%
and 
\begin{equation}
\Delta H_{n}\equiv \sum_{n^{\prime }}M_{nn^{\prime }}^{-1}\sum_{n^{\prime \prime
}}\frac{\Gamma _{n^{\prime }n^{\prime \prime }}}{F_{n^{\prime }}} .
\label{21}
\end{equation}%
We see that the effective magnetic field (\ref{star})
acting on an electron is given by 
\begin{equation}
h_{eff}=hH_{n}+Um(H_{n}+\Delta H_n).
\end{equation}%
It is interesting to observe that the dynamics of the two terms is {\em %
different}, i.e., the correlation problem leads to an asymmetry between the
external ($h$) and the induced ($Um$) effective magnetic fields. The origin
of this asymmetry lies in the self-consistency equation (\ref{3}), where $h$
enters through $\delta G_{\sigma n}$, while $Um$ enters both through $\delta
G_{\sigma n}$ and $\Sigma _{\sigma n}$. In the Hartree-Fock approximation,
the frequency dependent factors $H_{n}$ and $\Delta H_{n}$ reduce to unity and zero, 
respectively, such
that $h_{eff}=h+Um$ becomes a {\em static} effective magnetic field.

We are now able to calculate the magnetic susceptibility $\chi ^{DMFT}(T,U)$. 
Noting that the magnetization $m$, Eq.(2), has nonvanishing contributions
only from $\delta G_{n\sigma }$, and using Eq.(\ref{13}) one finds 
\begin{equation}
m=-\frac{1}{\beta }\sum_{n}H_{n}F_{n}h-\frac{1}{\beta }%
\sum_{n}(H_{n}+\Delta H_n)F_{n}Um.
\end{equation}%
The linear magnetic susceptibility is then obtained from $m=\chi
^{DMFT}(T,U)h$ as%
\begin{equation}
\chi ^{DMFT}(T,U)=\frac{\chi _{0}(T,U)}{1-U\left[\chi _{0}(T,U)+\Delta \chi_0(T,U)\right]},
\label{14}
\end{equation}%
where 
\begin{equation}
\chi _{0}(T,U)=-\frac{1}{\beta }\sum_{n}H_{n}F_{n},  \label{15}
\end{equation}%
and 
\begin{equation}
\Delta \chi _{0}(T,U)=-\frac{1}{\beta }\sum_{n}\Delta H_{n}F_{n}.  \label{15bis}
\end{equation}%
Eq. (\ref{14}) with Eqs. (\ref{15}) and (\ref{15bis}) is one of the main
results of our paper. We note that the expression for the static
susceptibility $\chi ^{DMFT}(T,U)$ in Eq. (\ref{14}) is deceptively simple.
Indeed, it has the {\em form} of the corresponding RPA expression, with the
Pauli susceptibility $\chi _{0}(T)$ of the non-interacting system replaced
by the susceptibilities $\chi _{0}(T,U)$ and $\Delta \chi _{0}(T,U)$
of the interacting system. If both $\tilde{\Sigma}_{n}$ and $\Gamma
_{nn^{\prime }}$ were neglected we would recover the well-known
Hartree-Fock result. The result for the static susceptibility can be
expressed in the RPA-like form 
\begin{equation}
\chi ^{DMFT}(T,U)=\frac{\chi(T,U)}{1-U\chi(T,U)}
\end{equation}%
with 
\begin{equation}
\chi(T,U)=\frac{\chi_{0}(T,U)}{1+U\Delta \chi%
_{0}(T,U)}.
\end{equation}

This equation suggests that, as in RPA, the susceptibility can be written as
an infinite series of bubble diagrams with (3-leg) vertex corrections. So far
it was not possible to deconvolute the corresponding
Bethe-Salpeter equation for the vertex corrections, since 
the scattering function (4-leg vertex), although is ${\bf k}$-independent in
the infinite dimensional theory, is still a complicated 
function of frequency.
It should be noted, however, that our algebraic derivation of the static
susceptibility is non-perturbative  anyway since it is not based on 
 direct diagrammatic resummations.

\section{Discussion}

The transition point between the paramagnetic and ferromagnetic phase is
determined by the divergence of the static susceptibility $\chi ^{DMFT}(T,U)$
in Eq. (\ref{14})%
\begin{equation}
1-U\left[\chi _{0}(T_{c},U)+\Delta\chi _{0}(T_{c},U) \right]=0.  \label{16}
\end{equation}%
Using the spectral representation for the summation over the Matsubara
frequencies one can write (\ref{16}) in a closed form as 
\begin{widetext}
\begin{equation}
1-\int_{-\infty }^{\infty }d\epsilon \left( \frac{1}{e^{\beta _{c}\epsilon
}+1}\right) \left\{ -\frac{1}{\pi }
Im\left[H(\epsilon +i0^{+})+\Delta H(\epsilon +i0^{+})\right] \right\} =0,
\label{18}
\end{equation}%
\end{widetext}
where $H(\epsilon +i0^{+})$ and $\Delta H(\epsilon +i0^{+})$
are  obtained by analytic continuation: $%
H_{n}\equiv H(i\omega _{n})\rightarrow H(\epsilon +i0^{+})$ and 
$\Delta H_{n}\equiv \Delta H(i\omega _{n})\rightarrow \Delta H(\epsilon +i0^{+})$.
It should be noted that $H(\epsilon )$ and $\Delta H(\epsilon )$
are still functions of
temperature because of the internal summations over Matsubara frequencies in
Eqs. (\ref{20}) and (\ref{21}). For given $U$ Eq. (\ref{18}) determines the Curie
temperature $T_{c}(U)$. Similarly, one may fix the temperature to determine
the critical interaction strength $U_{c}(T)$. In this respect Eqs. (\ref{16}), (\ref{18}) 
are generalizations of the Stoner criterion \cite{stoner33} $%
U_{c}^{Stoner}=1/N^{0}(\epsilon _{F})$ obtained in Hartee-Fock theory. In
general Eq. (\ref{18}) implies that, due to the inclusion of genuine
correlations, the transition point to the ferromagnetic phase is not merely
determined by the density of states {\em at} the Fermi level but, rather, by
the density of states at {\em all} energies. Due to the increase in the
kinetic energy the value of $U_{c}$ is reduced by an asymmetric DOS,
especially if the DOS has a singularity at the lower band edge. This had
already been found in the numerical solution of the DMFT equations for a
model DOS using Monte-Carlo simulations \cite{wahle98,ulmke98} and in the
approximate treatment of DMFT within the modified perturbation theory \cite%
{nolting00}. Eq. (\ref{18}) confirms these numerical findings analytically
and explains the origin as a correlation effect.

Expanding the static susceptibility (\ref{14}) around $T_{c}(U)$ (for $%
T>T_{c}>0$) we find a Curie-Weiss law 
\begin{equation}
\chi ^{DMFT}(T,U)=\frac{C^{DMFT}(T_{c}(U),U)}{T-T_{c}(U)},  \label{19}
\end{equation}%
with 
\begin{equation}
C^{DMFT}(T_{c}(U),U)=-\frac{\chi _{0}(T_{c}(U),U)}{
U\left\{\frac{d\left[\chi _{0}(T,U)+ \Delta \chi _{0}(T,U)\right]}{
 dT}\right\}_{T_{c}(U)}},
\end{equation}
which is hence seen to be a genuine property of the DMFT. The critical
exponent $\gamma =1$ is in accordance with the mean-field nature of the DMFT,
which neglects short-range spatial correlations between the electrons. 

In the similar manner one can show that at $T=0$, where the transition becomes
a quantum phase transition,
the static spin susceptibility diverges as the control parameter $U$ approaches
$U_c(T=0)$ from below
\begin{equation}
\chi^{DMFT}(T=0,U) \sim \frac{1}{U_c(T=0)-U},
\end{equation} 
with the mean field exponent $ \gamma=1$.

While the dynamics, i.e., the effect of temporal correlations, is found to
be very important in determining non-universal quantities such as the
critical temperature or the Curie constant, we conclude that it apparently does not affect
the universal scaling properties of the paramagnetic-to-ferromagnetic phase
transition within the DMFT. \bigskip 

\begin{acknowledgments}
We thank Dr. Matthias Vojta for very useful discussions. KB also thanks
Prof. V. Jani\v{s} and Dr. R. Bulla for important comments.  
This work was
supported by a Fellowship of the Alexander von Humboldt-Foundation (KB), and
through SFB 484 of the Deutsche Forschungsgemeinschaft.
\end{acknowledgments}


\begin{references}


\bibitem{georges96} A. Georges, G. Kotliar, W. Krauth, M. J. Rozenberg, Rev.
Mod. Phys. {\bf 68}, 13 (1996).

\bibitem{pruschke95} Th. Pruschke, M. Jarrell, J.K. Freerick, Adv. in Phys. 
{\bf 44}, 187 (1995).

\bibitem{voll93} D. Vollhardt, {\em Correlated Electron Systems},
ed. V. J. Emery, (World-Scientific, Singapore, 1993), p. 57.

\bibitem{metzner89} W. Metzner, D. Vollhardt, Phys. Rev. Lett. {\bf 62}, 324 (1989).

\bibitem{fazekas99} P. Fazekas, {\em Lecture Notes on Electron Correlation
and Magnetism} (World Scientific, Singapore 1999).

\bibitem{voll00} D. Vollhardt, N. Bl\"umer, K. Held, M. Kollar,
in {\em Band-Ferromagnetism}, eds. K. Baberschke, et al., Lecture Notes in Physics, 
vol. 580 (Springer, Heidelberg, 2001), p. 191.

\bibitem{jarrell} M. Jarrell, Phys. Rev. Lett. {\bf 69}, 168 (1992); M. Jarrell, Th. Pruschke,
Z. Phys. {\bf B 90}, 1891 (1993).

\bibitem{ulmke98} M. Ulmke, Eur. Phys. J. {\bf B 1}, 301 (1998).

\bibitem{jarrellbis} M. Jarrell, H. Akhlaghpar, Th. Pruschke, {\em Quantum Monte Carlo Method
in Condensed Matter Physics}, ed. M. Suzuki, (World-Scientific, Singapore, 1993), p.221.

\bibitem{vojta} We are grateful to M. Vojta for an illuminating discussion
of this point.

\bibitem{belitz99} D. Belitz, T. R. Kirkpatrick, T. Vojta, Phys. Rev. Lett. {\bf 82}, 
4707 (1999).




\bibitem{wolf} E. P. Wohlfarth, Philos. Mag. {\bf 42}, 374 (1951); Rev. Mod. Phys. {\bf 25}, 211 (1953);
M. Shimizu, Rep. Prog. Phys. {\bf 44}, 329 (1981); T. Moriya, {\em Spin Fluctuations in 
Itenerant electron Magnetism}, (Springer Verlag -  Heidelberg, 1985).

\bibitem{janis} For a discussion see, for example, V. Jani\v{s}, J. Ma\v{s}ek, D. Vollhardt,
Z. Phys. {\bf B 91}, 325 (1993). 

\bibitem{wahle98} J. Wahle, N. Bl\"umer, J. Schlipf, K. Held, D. Vollhardt,
Phys. Rev. {\bf B 58}, 12749 (1998).

\bibitem{stoner33} E. C. Stoner, Phil. Mag. {\bf 15}, 1018 (1933).

\bibitem{nolting00} W. Nolting, {\em Lectures on the physics of highly
 correlated electron systems IV}, pp. 118-225, ed. F. Mancini (AIP, 2000).


\end{references}

\end{document}